\documentclass[twocolumn,showpacs,superscriptaddress,prl]{revtex4}

\usepackage{graphicx}%
\usepackage{dcolumn}
\usepackage{amsmath}

\newcommand {\dedx} {$\langle dE/dx_{res} \rangle$}
 
\makeatletter
\def\btt#1{\texttt{\@backslashchar#1}}%
\DeclareRobustCommand\bblash{\btt{\@backslashchar}}%
\makeatother

\begin{document}


\title{New Measurement of the Cosmic-Ray
Positron Fraction from 5 to 15 GeV}

\author{J.J.~Beatty}
\thanks{Now at
Department of Physics, Ohio State University, Columbus, OH 43210}
\affiliation{Department of Physics,
Pennsylvania State University, University Park, PA 16802}
\author{A.~Bhattacharyya}
\affiliation{Department of Physics, Indiana University, Bloomington, IN 47405}
\author{C.~Bower}
\affiliation{Department of Physics, Indiana University, Bloomington, IN 47405}
\author{S.~Coutu}
\affiliation{Department of Physics,
Pennsylvania State University, University Park, PA 16802}
\author{M.A.~DuVernois}
\thanks{Now at School of Physics and Astronomy,
University of Minnesota, Minneapolis, MN 55455}
\affiliation{Department of Physics,
Pennsylvania State University, University Park, PA 16802}
\author{S.~McKee}
\affiliation{Department of Physics, University of Michigan, Ann Arbor, MI 48109}
\author{S.A. Minnick}
\thanks{Now at Department of Physics, 
Kent State University Tuscarawas Campus, New Philadelphia, OH 44663}
\affiliation{Department of Physics,
Pennsylvania State University, University Park, PA 16802}
\author{D.~M\"uller}
\affiliation{Enrico Fermi Institute and Department of Physics,
University of Chicago, Chicago, IL 60637}
\author{J.~Musser}
\affiliation{Department of Physics, Indiana University, Bloomington, IN 47405}
\author{S.~Nutter}
\affiliation{Department of Physics \& Geology,
Northern Kentucky University,
Highland Heights, KY 41099}
\author{A.W.~Labrador}
\thanks{Now at
California Institute of Technology, Space Radiation Laboratory,
Pasadena, CA 91125}
\affiliation{Enrico Fermi Institute and Department of Physics,
University of Chicago, Chicago, IL 60637}
\author{M.~Schubnell}
\thanks{Corresponding author.\\Email address: schubnel@umich.edu}
\affiliation{Department of Physics, University of Michigan, Ann Arbor, MI 48109}
\author{S.~Swordy}
\affiliation{Enrico Fermi Institute and Department of Physics,
University of Chicago, Chicago, IL 60637}
\author{G.~Tarl\'e}
\affiliation{
Department of Physics, University of Michigan, Ann Arbor, MI 48109}
\author{A.~Tomasch}
\affiliation{Department of Physics, University of Michigan, Ann Arbor, MI 48109}

\date{\today}

\begin{abstract}
We present a new measurement of the cosmic-ray positron fraction at
energies between 5 and 15 GeV with the balloon-borne HEAT-pbar
instrument in the spring of 2000.
The data presented here are compatible with our previous measurements,
obtained with a different instrument. The combined data from the three
HEAT flights indicate a small positron flux of non-standard origin above 5 GeV.
We compare the new measurement with earlier data obtained with the
HEAT-$e^\pm$ instrument, during the opposite epoch of the solar cycle,
and conclude that our measurements do not support predictions of charge
sign dependent solar modulation of the positron abundance at 5 GeV.
\end{abstract}

\maketitle

Space-borne and high altitude balloon experiments have
collected a considerable amount of data on the all-electron component
in the cosmic radiation.
Between several MeV and about 50 GeV the energy spectra of electrons
and positrons have been observed separately with sufficient precision
to permit comparison with production and propagation models.

The relative abundances of positive and negative electrons indicate
that the majority of the electron component $(e^\pm)$ consists
of negative electrons. They are thought to be accelerated in the
same Galactic sources that also generate the nuclear cosmic rays,
but their observed energy spectrum is strongly affected by
synchrotron and inverse Compton energy losses during propagation.

An additional component of electrons and positrons in nearly equal
proportion amounts to about 10\% of the total electron intensity,
and is attributed to the decay of secondary particles (mostly pions)
generated in hadronic interactions of cosmic-ray nuclei in the
interstellar medium. 
These positrons constitute only a small fraction ($<$ 0.5\%) of the
total observed cosmic-ray intensity, yet if they are purely secondaries
they can be used as an effective probe of cosmic-ray propagation
through the Galaxy. Their fraction is then expected to decline slowly
with energy because of the declining path length of the primary
nuclei at high rigidities.

Recent observations \cite{barwick95, barwick97apj, boezio2000, ams_posi}
confirm the almost exclusively secondary nature of positrons up
to a few GeV.
However, a possible structure in the positron fraction near 8 GeV has been
observed with the HEAT-e$^\pm$ instrument \cite{barwick97apj, coutu} which
may defy a simple explanation.

It has been suggested that a small positron component could originate
from particle interactions in nearby astrophysical
sources \cite{zhang, chi, harding, dogiel} or may be generated
through the annihilation of dark matter particles in the Galactic
halo \cite{tylka89,turner90,kt91,jungman96,kane2002,kane2002b,baltz2002, baltz2003, hooper, eichler}.
Such a primary positron component could lead to observable features
such as those indicated by the HEAT-e$^\pm$ measurements.

A new version of the HEAT instrument, HEAT-pbar was designed to
observe the high-energy cosmic-ray antiproton flux but it is also
suited for the observation of electrons and positrons at energies
below $\approx$ 15 GeV. The instrument utilizes a multiple dE/dx vs.~rigidity
technique to identify cosmic-ray particles by mass and charge.
The ionization loss of relativistic particles is sampled in two
stacks, each of 70 multiwire proportional chambers, filled with
a Xe/CH$_4$ mixture. They are located above and below a central
superconducting magnet spectrometer which measures the particle's
rigidity and charge sign \cite{barwick97nim}.
Scintillators at the top and bottom of the detector system measure
the time of flight and, together with a scintillator just above the
spectrometer, form the trigger. Unambiguous discrimination between
protons, antiprotons, positrons,
electrons, $\pi^{+}$ and $\mu^{+}$, $\pi^-$ and $\mu^{-}$, is achieved \cite{beach2001}.
This instrument was launched from Ft.~Sumner, NM on June 3, 2000 and
was at float altitude for 22 h, at an average atmospheric
overburden of 7.2 g/cm$^2$. The vertical geomagnetic rigidity
cut-off along the flight path varied little and averaged 4.2 GV.

The time of flight (ToF) system measures the particle velocity
$\beta = v/c$ with a resolution of $\sigma_\beta$ = 0.09, permitting
complete rejection of upward-going particles. The scintillator signals
measure the magnitude of the particle's electric charge and
select singly charged particles, with
a resolution of $\sigma_Z$ = 0.14 (in charge units) for each counter.

The particle's sign of charge and rigidity R=pc/Ze are determined with
the magnet spectrometer which has a field of about one Tesla.
We retain only events with at least 13 usable tracking
points in the magnet's bending plane (out of a total of 17) and at
least 6 points in the non-bending plane (out of a total of 8).

Only the smaller 50\% of the ionization signals in the 140 chambers
of the multiple dE/dx system are retained to form the restricted
average signal  $\langle dE/dx_{res} \rangle$.
Owing to higher electromagnetic energy losses, e$^\pm$ induced events
tend to result in a greater number of hits in the ionization chambers
of the dE/dx system than heavier singly charged particles. Therefore
we only select events with a large total number (64 or more) of
channels used in calculating \dedx \space and we require a minimum
number of hits in the entire dE/dx stack (155).

\begin{figure} [th]
\includegraphics[width=0.84\linewidth]{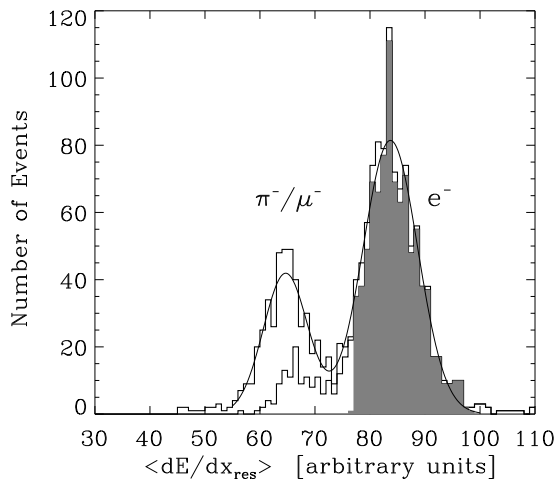}
\caption{Distribution of the average restricted energy loss for particles
in the rigidity range 4.5 -- 6.0 GV identified as negatively
charged. The Gaussian functions were fitted prior to the
final event selection in order to obtain a well-defined association
between average restricted energy loss and particle species. After the
final event selection is applied, non-electron events are
significantly suppressed (lower histogram). All particles within the
shaded area of the histogram, determined as described in the
text, are selected as electrons (and as positrons in the corresponding
positive rigidity bin).
}
\label{figure_1a}
\end{figure}

\begin{figure} [h]
\includegraphics[width=0.84\linewidth]{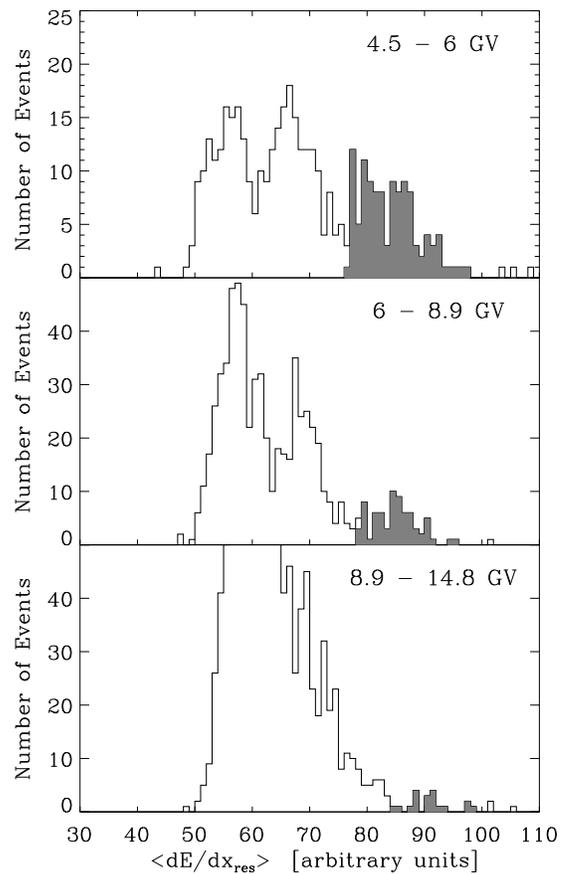}
\caption{Distributions of average restricted energy loss for events
  after all selection criteria have been applied. The
three rigidity bands, 4.5 -- 6.0 GV, 6.0 -- 8.9 GV,
and 8.9 -- 14.8 GV, are shown from top to bottom, respectively. 
We accept events under the shaded area as
positrons.
The strong relativistic rise in the energy loss for hadrons
and mesons compared to electrons (which are already heavily
relativistic at
GeV energies) results in the mass peak of the particle species moving
closer together with increasing rigidity and ultimately limiting the
particle identification. 
}
\label{figure_1b}
\end{figure}

Distributions of  \dedx \space for particle tracks having satisfied
all of the criteria are produced in three rigidity bands
(4.5 -- 6.0, 6.0 -- 8.9, and 8.9 -- 14.8 GV). To illustrate the
particle separation power of the instrument, the  \dedx \space
distribution for negative particles with rigidities between 4.5 GV and
6.0 GV is shown in Fig.~\ref{figure_1a}.
Electrons are clearly separated from muons and pions
whose small mass difference can not be distinguished by our instrument.  
Each of the three negative \dedx \space  distributions
is fitted with a sum of Gaussian functions and the mean and
standard deviation for the  $\pi^-/\mu^-$ mass peak (m$_{\pi\mu}$,
$\sigma_{\pi\mu}$) and the electron mass peak (m$_{e}$, $\sigma_{e}$)
are determined in each of the three rigidity bins.
We define a particle as an electron (or positron) if its
restricted average dE/dx signal falls between (m$_{\pi\mu}$+3$\sigma_{\pi\mu}$)
and (m$_{e}$+3$\sigma_{e}$) in the respective rigidity bin.
To further improve the discrimination against protons and mesons
we group the 140 proportional chambers into ten modules and
require that the average measured energy losses in each 
module exceed a given threshold. This selection strongly suppresses
protons and mesons while it retains almost all electrons
(see Fig.~\ref{figure_1a}). 
The residual contamination due to proton/muon/pion spillover was
obtained by fitting and extrapolating the p/$\mu$/$\pi$ distributions
in Fig.~\ref{figure_1b}. The background thus estimated for the
three rigidity bins are 0.4, 0.1, 1.0 events, respectively.
Note that the selection criteria
employed here were tuned to optimize the identification and
statistical significance of the e$^\pm$ peaks, and therefore
differ from those employed in the antiproton analysis of the
same data set \cite{beach2001}.

The atmospheric overburden during the HEAT-pbar flight varied between
4.5 and 11 g/cm$^2$, for an average overburden of 7.2 g/cm$^2$,
which resulted in significant numbers of atmospheric secondaries.
A Monte Carlo simulation, described elsewhere \cite{barwick98}, is
used to obtain the correction factors for atmospheric positron and
electron production.
The corrections for the positrons vary between
44\% and 52\%, depending on energy, and those for the electrons
between 5\% and 6.3\%.
Uncertainties in these atmospheric corrections result in a systematic
uncertainty of about $\pm$0.01 on the positron fraction.

\begin{table} [b]
\caption{Summary of event selection results and the calculated
positron fraction (in
10$^{-2}$). E$_{ToA}$ is the particle kinetic energy at the top of
the atmosphere. N$_{e^+}$ and N$_{e^-}$ are the number of observed
positrons and electrons for each energy bin, respectively.
N$_{e^+}^{cor}$ and N$_{e^-}^{cor}$ are the extrapolated number
 of positrons and electrons at the top of the atmosphere.}
\begin{ruledtabular}
\begin{tabular}{lrrrrc}
\multicolumn{1}{c}{E$_{ToA}$[GeV]}
& \multicolumn{1}{c} {N$_{e^+}$}
& \multicolumn{1}{c} {N$_{e^-}$}
& \multicolumn{1}{c} {N$_{e^+}^{cor}$}
& \multicolumn{1}{c} {N$_{e^-}^{cor}$}
& \multicolumn{1}{c} {$e^+/(e^+ + e^-$)[$\times 10^2$]}\\
\hline
5.0 -- 6.7 & 112 & 902 & 55 & 845 & 6.2 $\pm$ 0.6\\
6.7 -- 9.9 &  71 & 712 & 40 & 673 & 5.5 $\pm$ 0.7\\
9.9 -- 16.4&  18 & 238 &  8 & 226 & 3.6 $\pm$ 0.9\\
\end{tabular}
\end{ruledtabular}
\label{positron_results}
\end{table}

The raw e$^\pm$ particle counts are obtained by tallying the events
with a restricted average dE/dx signal that falls within a selected
region as described above and indicated in Fig.~\ref{figure_1b}. 
These counts are summarized in
Table \ref{positron_results} (N$_{e^+}$ and N$_{e^-}$.)
They are corrected for the atmospheric secondaries, and the
positron fractions calculated.

Finally, a correction factor of 1.109 is applied to rigidities at the
instrument, to account on average for radiative energy losses by
the electrons and positrons,
and to correct to the top of the
atmosphere.
The correction is done in an average way, rather than on an
event-by-event basis, because the events are grouped in narrow
rigidity bins
at the instrument prior to their identification as electrons or
positrons based on their energy losses.
This factor is calculated \cite{barwick98} as
$\alpha^{t/\alpha ln 2}$, where $t$ is the average atmospheric
depth in radiation lengths and $\alpha$ = 3.1 is taken for the spectral
index of the primary electron flux.

\begin{figure}[th]
\includegraphics[width=0.82\linewidth]{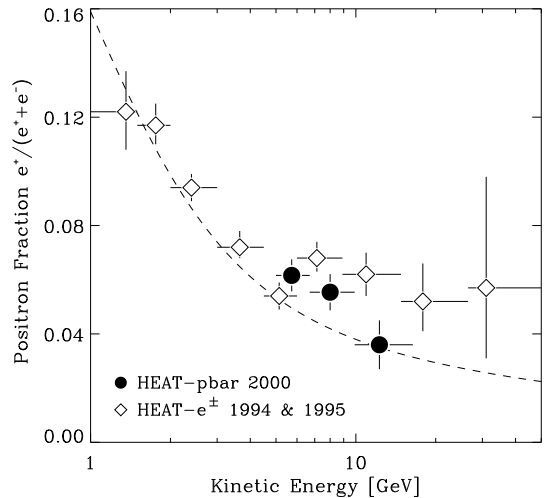}
\caption {The positron fraction measured by HEAT-pbar compared to
measurements with a different instrument (HEAT-e$^\pm$
\cite{barwick97apj});
The curve shows the expectation
from the model calculation in \cite{ms98}.
}
\label{figure_2}
\end{figure}

\begin{figure}[t]
\includegraphics[width=0.95\linewidth]{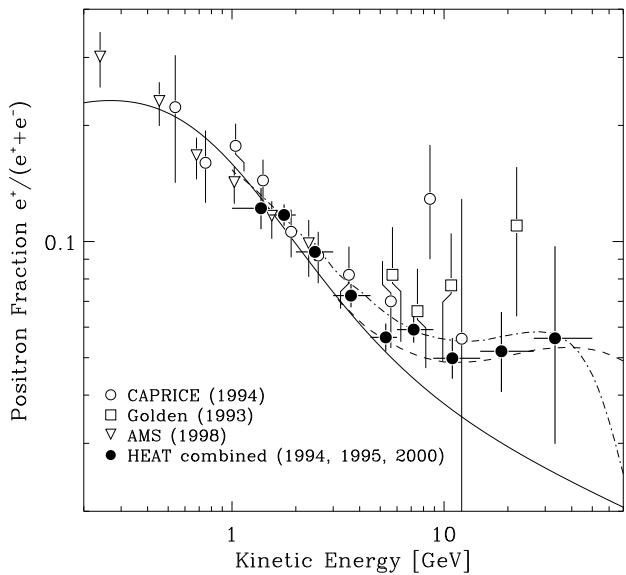}
\caption{
The positron fraction as a function of energy for the combined
HEAT-$e^\pm$ and HEAT-pbar data
, compared to model predictions and other recent measurements (
CAPRICE \cite{boezio2000},
Golden \cite {golden96},
AMS \cite{ams_posi}). Dates in parentheses give the year of
the measurement and not the publication.
The solid curve is the positron
fraction based on a purely secondary production of positrons
given by \cite{ms98}. The dashed and dot dashed curves are the ratios
including contributions from Higgsino LSP decay \cite{kane2002b} and
gamma-ray pulsars \cite{zhang},
respectively.
}
\label{figure_3}
\end{figure}

Fig.~\ref{figure_2} shows the final corrected positron
fractions, $e^+/(e^+ + e^-$),  as a function of energy, compared with
the previous combined HEAT-e$^\pm$ results \cite{barwick97apj} from
the 1994 and 1995 flights. The accessible energy range for our
measurements is limited towards the lower energies by the geomagnetic
cut-off and at high energies by the particle separation capability of
the instrument. The result presented here is in agreement with the
HEAT-e$^\pm$ measurements obtained with a very different technique.
The result is also compatible with the predicted
ratio from \cite{ms98} which assumes that all positrons are of
secondary origin (dashed curve in Fig.~\ref{figure_2}).
Although the new data set weakens the case for a feature at $\approx$
8 GeV, an excess of positrons in this energy range cannot be ruled out.
The combined dataset of the HEAT-$e^\pm$ and HEAT-pbar experiments is
shown in Fig.~\ref{figure_3} along with a compilation of recent
measurements and predicted flux ratios.

We compare our data to a computation of the cosmic-ray secondary
positron (and electron) spectrum in a diffusive model for
Galactic cosmic-ray propagation provided by \cite{ms98} for
different source injection spectra. With injection spectra based
on the locally measured nuclear cosmic-ray spectra,
the data follow the general trend of the model prediction up
to a few GeV, beyond which the observed positron fraction is
higher than the calculated one. This enhancement is plausible
in a variety of different theoretical scenarios.

Motivated by a measured diffuse gamma-ray spectrum above 1 GeV
that is much harder than expected \cite{hunter97}, local
cosmic-ray fluxes have been calculated based on hard proton
injection spectra \cite{ms98}.
Interestingly, this does not only lead to a better fit of the
diffuse gamma-ray data but also to an enhancement in the
positron fraction above a few GeV. However, measurements of
cosmic-ray antiprotons provide rather substantial evidence
against the idea of explaining the diffuse gamma-ray excess with
a hard nucleon spectrum \cite{beach2001}.
Therefore, if future measurements confirm the positron excess,
then there may be additional sources for the cosmic-ray positrons.

An interesting suggestion is that annihilating dark matter particles
in the Galactic halo, perhaps supersymmetric particles, are a
source of high-energy positrons\cite{tylka89,turner90,kt91,jungman96,kane2002,kane2002b,baltz2002, baltz2003, hooper, eichler}, 
and references therein). In conventional models, the lightest
supersymmetric particle (e.g., neutralino) is assumed to have a mass
larger than the W mass and direct decay will produce a substantial
positron signal at about half the neutralino mass. Subsequent heavy lepton
and b decays could result in an enhancement of the positron fraction
at much lower energies, as shown
in Fig. \ref{figure_3} \cite{kane2002b}.

Other possible contributions to the cosmic-ray positron flux have been
proposed, including synchrotron produced e$^\pm$ pairs from Galactic
pulsars \cite{zhang}. Whether the apparent excess in the HEAT data
is caused by one of those processes
is not clear and
can only be determined if further positron measurements provide
improved statistical accuracy, and extend to higher energies.

It has been suggested that at energies below a few GeV
the measured positron fraction may reflect charge sign dependent
effects of the solar modulation \cite{clem1996, clem2002} and that
discrepancies between results from different experiments can be
related to this modulation effect. The most recent solar polar
field reversal occurred just prior
to the HEAT-pbar measurement and thus data from the three HEAT flights
can be used to investigate this proposed effect. In the energy range
5--6 GeV - the lowest energy common to all observations - the measured
positron fraction is consistent with what we measure
in 1994 and 1995 with HEAT-$e^\pm$ (see Fig. \ref{figure_2}).
At 5 GeV, Clem et al. \cite{clem1996} predict a decrease in the
positron fraction by about 40\%, which is inconsistent with our
measurements. It would be of interest to reconcile our result with the
measured dependence of the antiproton flux on solar activity by the
BESS group \cite{bess1, bess2}.

In conclusion, we find that a primary contribution to the positron
intensity above a few GeV can still not be ruled out.

\begin{acknowledgments}
This work was supported by NASA grants No. NAG 5-5058,
No. NAG 5-5220, No. NAG 5-5223, and No. NAG 5-5230.
We wish to thank the National Scientific Balloon Facility and the NSBF
launch crews for their excellent support of balloon missions and
we gratefully acknowledge contributions from D. Kouba, M. Gebhard, S. Ahmed,
and P. Allison. We also thank S. Beach for essential support of the mission. 
\end{acknowledgments}

\bibliography{posi2004}

\end{document}